\documentstyle[12pt]{article}

\newlength{\extraspace}
\newlength{\extraspaces}
\def\bsklength{.8mm} %{2mm} % for more than double spacing
\newcommand{\beq}{\begin{equation}}
\newcommand{\eeq}{\end{equation}}

\newcommand{\beqa}{\begin{eqnarray}}
\newcommand{\eeqa}{\end{eqnarray}}

\newcommand{\newsection}[1]{
\vspace{6mm}
\pagebreak[3]
\addtocounter{section}{1}
\setcounter{equation}{0}
\setcounter{subsection}{0}
\noindent{\large \bf \thesection. #1}
\nopagebreak
\medskip
\nopagebreak}

\def\is{\!\!\!\!=\!\!\!\!}

\def\half{{\textstyle{1\over 2}}}
\def\pa{\partial}

\def\thetabar{{\overline{\theta}}}

\def\CD{{\cal D}}
\def\CE{{\cal E}}

\def\CL{{\cal L}}

\def\CO{{\cal O}}

\renewcommand{\hat}{\widehat}
\renewcommand{\tilde}{\widetilde}
\renewcommand{\det}{{\rm det}}
\renewcommand{\bar}{\overline}

\newcommand{\e}{{\rm e}}

\def\d{{\rm d}}

\newcommand{\pth}{{\pa\ \over\pa\theta}}
\newcommand{\pthb}{{\pa\ \over\pa\bar\theta}}

\begin{document}
\setcounter{page}{0}
\addtolength{\baselineskip}{\bsklength}
\thispagestyle{empty}
\renewcommand{\thefootnote}{\fnsymbol{footnote}}	%for symbols

\begin{flushright}
{\sc MIT-CTP-2624}\\
hep-th/9704196\\
April 1997
\end{flushright}
\vspace{.4cm}

\begin{center}
{\Large
{\bf{Supersymmetric Dilatations in the Presence of Dilaton}}}\\[1.2cm] %title
{\rm HoSeong La}%						%author
\footnote{e-mail address: hsla@mitlns.mit.edu\\			%email
This work is supported in part by funds provided by the U.S. Department of
Energy (D.O.E.) under cooperative research agreement \#DF-FC02-94ER40818.
 }\\[3mm]							
{\it Center for Theoretical Physics\\[1mm]		%address
Laboratory for Nuclear Science\\[1mm]
Massachusetts Institute of Technology\\[1mm]
77 Massachusetts Avenue\\[1mm]
Cambridge, MA 02139-4307, USA} \\[1.5cm]

%{\sc Abstract}\\[1cm]
{\parbox{14cm}{
\addtolength{\baselineskip}{\bsklength}
\noindent
The supersymmetric generalization of dilatations in the presence of the dilaton
is defined. This is done by defining the supersymmetric dilaton geometry which
is motivated by the supersymmetric volume preserving diffeomorphisms. The
resulting model is classical superconformal field theory with an additional
dilaton-axion supermultiplet coupled to the supersymmetric gauge theory,  where
the dilaton-axion couplings are nonrenormalizable.  The possibility of
spontaneous scale symmetry breaking is investigated in this context. There are
three different types of vacua with broken scale symmetry depending on the
details of the dilaton sector: unbroken supersymmetry, spontaneously broken
supersymmetry and softly broken supersymmetry. If the scale symmetry is broken
in the bosonic vacuum,  then the Poincar\'e supersymmetry must be broken at the
same time. If the scale symmetry is broken in the fermionic vacuum but the
bosonic vacuum remains invariant, then the Poincar\'e supersymmetry can be
preserved as long as the R-symmetry breaking is specifically related to the
scale symmetry breaking. 

\bigskip
PACS: 11.30 11.30.Ly 11.30.Pb 11.30.Qc
} 
}

%{Submitted to {\it Nuclear Physics B}} 

\end{center}
\noindent
\vfill

%%%%%%%%%%%%%%%%%%%%%%%%%%%%%%%%%%%%%%%%

\setcounter{section}{0}
\setcounter{equation}{0}
\setcounter{footnote}{0}
\renewcommand{\thefootnote}{\arabic{footnote}}	%for numbers
\newcounter{xxx}
\setlength{\parskip}{2mm}
\addtolength{\baselineskip}{\bsklength}
\newpage

\newsection{Introduction}

\noindent
The dilaton is a hypothetical particle which is a Goldstone boson associated
with spontaneous breaking of scale symmetry as an analog to the pion for chiral
symmetry breaking\cite{cole}. Scale symmetry is one of the most fascinating
concepts in physics despite the fact that nature does not seem to respect it at
all  at least in the macroscopic scale. Microscopically, since the MIT-SLAC
deep inelastic scattering  experiments\cite{dpinel} the role of dilatations in
physics has attracted  fair amount of attention. Knowing the fact that we live
in the world of a given scale, the classical scale
symmetry\cite{wess}\cite{rscale}\cite{anom}\cite{spon}  based on the
dilatations of local coordinates in a Lorentz frame is destined to be broken at
that given scale. Scale symmetry breaking in principle can be either explicit
or spontaneous. It turns out that in a simple model with a scalar field the
classical scale symmetry is anomalous at the quantum level, hence it is 
explicitly broken\cite{anom}. In more realistic cases like massless QED or
gauge theories, scale symmetry is also broken by the trace anomaly\cite{racd}.
This does not allow any room for the dilaton to be introduced. Nevertheless,
there have been attempts to introduce spontaneous breaking of scale
symmetry\cite{cole}. 

If we look into the matter more carefully, the existence of anomalies does not
necessarily rule out any possibility of spontaneous symmetry breaking.
If spontaneous symmetry breaking occurs in a sector different from the 
anomalous sector, it is possible. In fact, a good example is the case of the
axion. Despite the axial anomaly, we can obtain spontaneous breaking of the
Peccei-Quinn symmetry and that the axion can be generated in QCD\cite{raxion}.
This seems to be the case for the dilaton too.
Furthermore, the dilaton does not transform like a quasiprimary field under
dilatations, it does not follow the standard structure of the renormalization
group either. So, in the presence of the dilaton the renormalization group
argument does not rule out the possibility of spontaneous breaking of scale 
symmetry.  

The dilaton appears more commonly in the context of gravity,  although it often
lacks an unambiguous way of defining the dilaton. Not all scalar fields
appearing in gravitational models are dilatons, but the  only one related to
the spontaneous breaking of scale symmetry is the dilaton. For example, in
Kaluza-Klein approaches a compactification of $(n+1)$ down to $n$ dimensional
spacetime introduces a scalar field which behaves like a dilaton in $n$
dimensions. But this is not a definition of the dilaton but just a coincidence
because $(n+k)$-to-$n$ compactifications can lead to a set of scalar fields
satisfying nonlinear $\sigma$-models, none of which has the property of the
dilaton\cite{cremjul}. 
The reason we sometimes consider this scalar field as a dilaton is the
coincidence observed in the low energy effective supergravity models of string
theory. A low energy effective ten dimensional supergravity action of string
theory can be derived from an eleven dimensional supergravity model action via 
compactification on a circle
and the Kaluza-Klein scalar field appears as the   dilaton in
the Einstein frame derived from the string theoretical  dilaton\cite{wittdil}.
In string theory the dilaton is more precisely and rather unambiguously defined
as follows: The dilaton in string theory is the spacetime scalar field coupled
to the world-sheet curvature scalar in the world-sheet nonlinear $\sigma$-model
action\cite{rdefdil}. This is the only massless scalar field 
and the theory has classical scale symmetry, hence it has to
be the gravitational dilaton.  This leads to the Kaluza-Klein scalar field in
the Einstein frame, in which the metric is a combination of the stringy dilaton
and the spacetime metric in the string frame, that is, the spacetime metric
defined in the worldsheet nonlinear $\sigma$-model  action. Thus the dilaton
that appears in the low energy effective supergravity action is not really
independent from the trace of the graviton, which often leads to dilaton
fluctuations involving the trace of the graviton\cite{polch}\cite{lanelson}.

To further understand the role of scale symmetry in nature, we need to address
the origin of the low energy scale symmetry. It is commonly believed that the
scale symmetry in Minkowski space is an analog to the (rigid) Weyl symmetry in
curved spacetime\footnote{See \cite{bd}, for example.}. This however is in
some sense unsatisfactory because of lack of any direct relationship. This
is all right if they are two independent symmetries. 
However, it is most unlikely that they are
unrelated because classical symmetries should not depend on the strength of the
gravity, yet the rigid Weyl ceases to make sense when the gravity is turned
off. Furthermore, 
low energy scale transformations involve changes of local coordinates,
but Weyl transformations do not. We need a more direct connection between
properties in curved spacetime and those in a local Lorentz frame particularly
for any spacetime symmetries.  This becomes an important issue if gravitational
effects get stronger. Particularly, in string-motivated supersymmetric models
the dynamics of the dilaton is crucial to understand the structure of the
coupling constants and
supersymmetry in the low energy\cite{wittdil}\cite{ds}\cite{kaplo}\cite{nilles}.

As a matter of fact, the rigid Weyl transformations can be reproduced by
diffeomorphisms. Hence, in my previous paper\cite{mydil} I started only with
Diff (diffeomorphism) symmetry without referring to the Weyl symmetry, then
derived the scale symmetry in Minkowski spacetime.   Note that in any
dimensional spacetime scale invariance does not necessarily imply Weyl
invariance, although it often implies conformal invariance. This signals that
it would be better to understand scale symmetry as part of  Diff, and that it
could provide a natural explanation of the relation between scale symmetry and
conformal invariance. Diff decomposes into SDiff (volume-preserving
diffeomorphisms) and CDiff (conformal diffeomorphisms)\footnote{For details of
SDiff, see \cite{sdiffgr}.}. Since SDiff preserves a volume element,
dilatations are not part of SDiff. This is the crucial structure used in 
ref.\cite{mydil} and we shall generalize it to the supersymmetric case in this
paper.

Dilatations are defined in terms of local coordinates by
\beqa
\label{e1}
x &\to & \e^\alpha x ,\\
\label{e2}
\Phi_{[d]}(x) &\to & \e^{d\alpha}\Phi_{[d]}(\e^\alpha x),
\eeqa
where $d$ is the scale dimension (or the conformal weight). Eq.(\ref{e1})
suggests dilatations should be expressed as diffeomorphisms, although
eq.(\ref{e2}) is not a result of a diffeomorphism. We however find that
$\Phi_{[d]}$ can be expressed as a dilaton-dressed field and that eq.(\ref{e2})
indeed becomes a result of a diffeomorphism. This can be done only in the 
presence of the dilaton so that in this context the scale symmetry naturally 
incorporates the  dilaton. As an important result, scale invariance 
automatically guarantees conformal invariance because both are just part of 
Diff invariance.

The idea of defining dilatations in
the presence of the dilaton without referring to the Weyl transformation 
involves the diffeomorphism (Diff) symmetry of the dilaton geometry
defined by the metric $g_{\mu\nu} = \e^{2\kappa\phi}\eta_{\mu\nu}$, where 
$\phi \equiv {1\over 2\kappa n} \ln|g|$ in $n$-dimensional spacetime. One of
the advantages of this new proposal is that dilatations in Minkowski spacetime 
are naturally related to the symmetry in curved spacetime, while the
usual way of thinking that the low energy dilatations are related to the
Weyl geometry of curved spacetime actually lacks any direct relationship.
The dilaton geometry realizes the conformal symmetry in the presence of the 
dilaton as much as the conformal geometry does for the conformal symmetry 
without the dilaton. I expect that this could be the right framework to 
investigate the dilaton physics in the low energy limit of unified theories
that incorporates the dilaton, e.g. string theory.

This paper is organized as follows: In section two, the dilaton geometry is
defined. In section three, the $N=1$ superconformal vector fields are explicitly
derived, which are needed to generalize the dilaton geometry to the
supersymmetric case in section four. In section five, a superconformally
invariant effective lagrangian is derived for the supersymmetric dilaton
geometry. In section six the possibility of spontaneous breaking of the
scale symmetry in this context is examined. Finally, some perspectives are
discussed in the conclusions and one appendix that explains the basics of
volume-preserving Diff and its supersymmetric generalization are given.

\newsection{Dilaton Geometry and Conformal Symmetry}

\noindent
In curved spacetime, infinitesimal Weyl transformations are given by
\beq
\label{eweyl}
\delta g_{\mu\nu} = 2 \epsilon \rho (x) g_{\mu\nu},
\eeq
for arbitrary function $\rho(x)$.
$\rho(x)$ is constant for a global (or rigid) Weyl transformation. 
Usually in literatures this global Weyl transformation is regarded as the
analog to a scale transformation in Minkowski space, hence relating scale
symmetry to Weyl symmetry. Weyl transformations are 
independent from coordinate changes contrary to scale transformations.
Thus, by simply taking the flat limit Weyl symmetry does not naturally lead 
to the scale symmetry.

Spacetime transformations involving local coordinates are diffeomorphisms,
commonly known as general coordinate transformations.
Under Diff, fields transform according to 
\beq
\label{e3}
\left(T_{\mu_1\cdots\mu_p} + \delta T_{\mu_1\cdots\mu_p}\right)
\d x^{\mu_1}\cdots\d x ^{\mu_p} =
T_{\mu_1\cdots\mu_p}(x+\delta x) \d(x+\delta x)^{\mu_1}
\cdots\d(x+\delta x)^{\mu_p}.
\eeq
Then $\delta T_{\mu_1\mu_2\cdots\mu_p}$ is nothing but the Lie derivative
along $\delta x$. 
Diff acts on the metric, for $v\equiv \delta x$ in $n$ dimensions, as
\beq
\label{emetr}
\delta g_{\mu\nu} = \nabla_\mu v_\nu + \nabla_\nu v_\mu. 
\eeq
In particular, if $v^\mu$ is a conformal Killing vector, 
then $\delta g_{\mu\nu}$ takes the form of eq.(\ref{eweyl}) with 
$\rho(x) = {1\over n}\nabla_\mu v^\mu$
and these transformations are conformal diffeomorphisms.
Contrary to Weyl transformations, where $\rho(x)$ is arbitrary, such $v^\mu$ 
must exist in this case. If so, one can say CDiff is a special case of Weyl.
If $v^\mu$ exists for constant $\rho$, the scale transformation based on
the Weyl is also nothing but Diff. 

In the presence of the dilaton we can take advantage of the above to the most.
The dilaton geometry is defined by a metric in the form of
\beq
\label{e4}
g_{\mu\nu} = \e^{2\kappa\phi}\eta_{\mu\nu},
\eeq
where $\e^{n\kappa\phi} = \sqrt{g}$ in terms of $g\equiv |\det g_{\mu\nu}|$ and
$\kappa^{-1}$ is the dilaton scale. The effect of introducing an explicit 
scale parameter, $\kappa$, is to let the dilaton have mass dimension 
$(n-2)/2$, where $n$ is the dimension of spacetime. Note that $\kappa$  is not
really a free parameter because we can always rescale it by rescaling $\phi$.
As far as gravity is concerned, the natural choice of this scale is the Planck
scale. But, here, instead of doing that, we will fix it later at any
phenomenologically proper scale so that we can study the dilaton in an energy
scale much lower than the quantum gravity scale. Since $\kappa$ always appears
in combination with $\phi$, fixing $\kappa$ actually requires a nontrivial
dilaton vacuum expectation value.

As emphasized in ref.\cite{sdiffgr}, if $\phi$ does not transform like a scalar 
under Diff, but the transformation property under Diff is dictated by that of
the metric,  then eq.(\ref{emetr}) leads to
\beq
\label{e5}
\delta\e^{\kappa\phi} ={\textstyle{1\over n}}\e^{\kappa\phi}D_\mu v^\mu,
\eeq
where 
\beq
\label{eqder}
D_\mu \equiv \pa_\mu + n\kappa\pa_\mu\phi .
\eeq
In terms of $g$,  $D_\mu \equiv \pa_\mu +\pa_\mu\ln\sqrt{g}$ and
$\pa_\mu\left(\sqrt{g} v^\mu\right) = \sqrt{g} D_\mu v^\mu$. $D_\mu$ is the
same as the covariant derivative $\nabla_\mu$ only when it acts on a covariant
vector, but, in general, they are different. Eq.(\ref{e5}) shows that
eq.(\ref{e4}) is not to be considered as a conformal gauge fixing  condition
globally, but is a local  expression of a metric in terms of a non-global
function $\phi$. For example, a density is not global  because it depends on a
choice of local coordinates. If eq.(\ref{e4}) were the  conformal gauge fixing
condition, it would lead to $\delta\e^{2\kappa\phi} = v^\mu\pa_\mu
\e^{2\kappa\phi}$. Under SDiff, $\phi$ behaves like a constant to make pure
$\phi$ Lagrangians manifestly SDiff-invariant.

It is important to notice that
the consistency condition between eq.(\ref{e5}) and eq.(\ref{e4}) is
\beq
\label{ecnf}
{\textstyle{2\over n}}\eta_{\mu\nu}\pa_\alpha v^\alpha =
\eta_{\mu\alpha}\pa_\nu v^\alpha + \eta_{\alpha\nu}\pa_\mu v^\alpha.
\eeq
Hence, diffeomorphisms of eq.(\ref{e4}) appear as conformal transformations of
the flat spacetime. This shows that the dilaton geometry describes the conformal
geometry of the flat spacetime in the presence of the dilaton.

Under dilatations $v^\mu = \alpha x^\mu$, eq.(\ref{e1}), the dilaton transforms
for infinitesimal $\alpha$, as 
\beq
\label{e6}
\delta\phi = \alpha\left({\textstyle{1\over\kappa}} + x^\mu\pa_\mu\phi\right),
\eeq
which is nothing but the dilatation property given in ref.\cite{cole}.
Note that the dilaton is not a (quasi)primary field, since it does not
transform homogeneously. This distinguishes the dilaton from other fields.
This also shows that the dilatations of the dilaton are results of 
diffeomorphisms. Sometimes, it is useful to introduce a field redefinition 
\beq
\label{echd1}
\chi \equiv \e^{\kappa\phi}.
\eeq
Under Diff, $\chi$ transforms in a not-so-inspiring way, 
but, under dilatations
\beq
\label{echd2}
\delta\chi = \alpha \left(1 + x^\mu\pa_\mu\right)\chi.
\eeq
Thus, although $\chi$ is not a scalar, it transforms like a scale-dimension-one
field. $\chi$ is mass-dimensionless.

To produce eq.(\ref{e2}) let us introduce a dilaton-dressed field $\Phi_{[d]}$
as
\beq
\label{e7}
\Phi_{[d]} \equiv \e^{d\kappa\phi}\Phi ,
\eeq
where $\Phi$ transforms like a scalar under Diff. This dilaton dressing does
not change the mass dimension of the field. Then under dilatations
\beq
\label{e8}
\delta\Phi_{[d]} = \alpha (d + x^\mu\pa_\mu)\Phi_{[d]}
\eeq
so that $\Phi_{[d]}$ is a (quasi)primary field\cite{mack}.
Such dressing is not needed for vector fields in four dimensions
because under dilatations
\beq
\label{e9}
\delta A_\mu = \alpha (1 + x^\lambda\pa_\lambda) A_\mu.
\eeq
This in particular leads to the YM term that does not couple directly to 
the dilaton in four dimensions, hence different from the Kaluza-Klein case.
In other than four dimensions vector fields still need dilaton dressing, leading
to direct YM-dilaton couplings.
Similarly, we can define all dimensionful fields in $n$ dimensions by 
properly dressing with the dilaton and the scale transformation properties
follow from the Diff. In this sense, the mass dimension of
a field is not necessarily the same as its scale dimension. For example, the
dilaton has mass dimension $(n-2)/2$, but its scale dimension is not even
defined.

One can also easily check that the dilaton is, after all, a Lorentz scalar, 
hence so is $\Phi_{[d]}$. Thus, from the low energy point of view $\Phi_{[d]}$
and $\phi$ are  indistinguishable from the usual scalar field. This clearly
shows that the dilatations in Minkowski space can be derived from the Diff of
virtual spacetime geometry of $g_{\mu\nu} = \e^{2\kappa\phi} \eta_{\mu\nu}$ and
we are never required to introduce Weyl symmetry.

From the supersymmetric point of view, the dilaton is inevitably associated
with the axion so that the dilaton, axion and dilatino form a supermultiplet.
In the dilaton geometry the axion can be easily incorporated by generalizing
$\chi$ to include a phase such that eq.(\ref{echd1}) is replaced by a 
new definition
\beq
\label{eax1}
\chi =\e^{\kappa\phi_c},\ \ \kappa\phi_c\equiv\kappa\phi +i {a\over f_a}
\eeq
and 
\beq
\label{eax2}
g_{\mu\nu} = \chi^*\chi\eta_{\mu\nu}.
\eeq 
Demanding that $\chi$ should still transform like a scale-dimension one field
under dilatations, the axion transformation rule can be obtained:
\beq
\label{eax3}
\delta a = \alpha x^\mu\pa_\mu a.
\eeq
The axion is a scale-dimensionless field. In general, under Diff the axion
transforms like a scalar so that
\beq
\label{eax3b}
\delta a = v^\mu\pa_\mu a.
\eeq

To make sure this field is really entitled to be named as the axion, we 
can in fact check the  relation between $a$ and the (gravitational) axion
associated with the antisymmetric field $B_{\mu\nu}$. In the dilaton geometry
the relation between the axion and $B_{\mu\nu}$ is given by
\beq
\label{eax4}
\pa_\sigma a = \half \e^{-2\kappa}\epsilon_{\sigma\lambda\mu\nu}\pa^\lambda
B^{\mu\nu}.
\eeq
From $\chi$ we can derive
\beq
\label{eax5}
\pa_\sigma a = {i\over 2}\e^{-2\kappa\phi}\left(
\chi\pa_\sigma\chi^* - \chi^*\pa_\sigma\chi\right).
\eeq
Thus, identifying
\beq
\label{eax6}
\epsilon_{\sigma\lambda\mu\nu}\pa^\lambda B^{\mu\nu} = 
i\left(\chi\pa_\sigma\chi^* - \chi^*\pa_\sigma\chi\right)
\eeq
the axion in eq.(\ref{eax1}) can indeed be identified as the standard axion.
We shall also later see that this axion is in fact the R-axion and
has the similar couplings as the axion
associated with the Peccei-Quinn symmetry.

For complex $\chi$ the dressing of the ordinary fields can be generalized in 
an obvious way except for fermions. Fermions should be dressed over each Weyl
components such as
\beq
\label{eax7}
\Psi_{[d]} \equiv {\psi_{1[d]\alpha}\choose \bar\psi_{2[d]}^{\dot\alpha}}
= \pmatrix{\chi^d& 0\cr 0& {\chi^*}^d}
{\psi_{1\alpha}\choose \bar\psi_{2}^{\dot\alpha}}
\equiv \pmatrix{\chi^d& 0\cr 0& {\chi^*}^d}\Psi .
\eeq

\newsection{Superconformal Vector Fields and Superconformal Algebra}

\noindent
Before I generalize the dilaton geometry to the supersymmetric case, let us
recapture the superconformal geometry. The superconformal geometry is described
by superconformal vector fields which satisfy the superconformal algebra.
The well known $N=1$ superconformal algebra is given 
by the follows\footnote{Our convention here follows that of \cite{WB}.}.

First, the usual conformal algebra
\beqa
\label{escf1a}
\left[M_{\mu\nu}, M_{\rho\sigma}\right] &\is& 
\eta_{\nu\rho}M_{\mu\sigma} -\eta_{\mu\rho}M_{\nu\sigma}
-\eta_{\nu\sigma}M_{\mu\rho} +\eta_{\mu\sigma}M_{\nu\rho},\\
\left[P_\mu, M_{\rho\sigma}\right] &\is& 
\eta_{\mu\rho}P_\sigma -\eta_{\mu\sigma}P_\rho,\\
\left[K_\mu, M_{\rho\sigma}\right]&\is& 
\eta_{\mu\rho}K_\sigma -\eta_{\mu\sigma}K_\rho,\\
\left[P_\mu, D_d\right] &\is& P_\mu,\\
\left[K_\mu, D_d\right] &\is& -K_\mu,\\
\left[P_\mu, K_\nu\right] &\is& 2\left(\eta_{\mu\nu}D_d - M_{\mu\nu}\right),
\eeqa
and for the ordinary Poincar\'e supersymmetry
\beqa
\label{escf1b}
\left[Q, M_{\mu\nu}\right] &\is& -\sigma_{\mu\nu} Q, \\
\left[Q, D_d\right] &\is& \half Q, \\ 
\left[Q, P_\mu\right] &\is& 0, \\ 
\left[Q, K_\mu\right] &\is& -\gamma_\mu S, \\ 
\{Q, \bar Q\} &\is& 2i\gamma^\mu P_\mu.
\eeqa
The superconformal part requires an additional S-supersymmetry
\beqa
\label{escf1c}
\left[S, M_{\mu\nu}\right] &\is& -\sigma_{\mu\nu} S,\\
\left[S, D_d\right] &\is& -\half S,\\
\left[S, P_\mu\right] &\is& \gamma_\mu Q,\\
\left[S, K_\mu\right] &\is& 0,\\
\{S, \bar S\} &\is& 2i\gamma^\mu K_\mu. 
\eeqa
The anti-commutator between Q- and S-supersymmetry
\beq
\label{escf1d}
\{S, \bar Q\} =
-2iD_d +2i\sigma^{\mu\nu}M_{\mu\nu} -3\gamma_5 R .
\eeq
As is well known, this shows the R-symmetry is inevitably required in the
superconformal symmetry.
Finally, those involving $R$,
\beqa
\label{escf1e}
\left[Q, R\right] &\is& i\gamma_5 Q, \\ 
\left[S, R\right] &\is& -i\gamma_5 S,
\eeqa
and all others vanish. We have written the algebra in terms of superconformal
vector fields so that the operator algebra follows by $\CO\to -i\CO$ 
prescription for any operator $\CO$ in the above except $Q$, $S$. 
This leads to the correct
eigenvalue problem for the hamiltonian $H=i\pa_0$. 

The superconformal vector fields that satisfy the above superconformal algebra 
can be derived by solving the superconformal condition
\beq
\label{econ1}
\left[D_\alpha, \hat{X}\right] = F_\alpha^{\ \beta}D_\beta,
\eeq
where $F_\alpha^{\ \beta}$ is a function in superspace. 
$\hat{X}$ is a vector field in superspace such that  
\beq
\label{econ2}
\hat{X} = V^\mu\pa_\mu 
+ \xi^\alpha{\pa\ \over\pa\theta^\alpha} 
+ \bar\xi_{\dot\alpha}{\pa\ \over\pa\bar{\theta}_{\dot\alpha}} .
\eeq
$V^\mu$ and $\xi$ are determined by eq.(\ref{econ1}) and we obtain
\beqa
\label{econ3}
V^\mu &\is& v^\mu 
+ i\left(\theta\sigma^\mu\bar\zeta - \zeta\sigma^\mu\bar\theta\right)
+{\textstyle{1\over 4}}\theta\theta
\left(\bar\sigma^\lambda\pa_\lambda\bar\zeta\sigma^\mu\bar\theta\right)
+\bar\theta\bar\theta\left(\theta\sigma^\mu\bar\sigma^\lambda\pa_\lambda\right)
+{\textstyle{1\over 4}}\theta\theta\bar\theta\bar\theta
\pa^\lambda\pa_\lambda v^\mu, \\
\xi^\alpha &\is& \zeta^\alpha +\theta\sigma^\mu\bar{\sigma}^\nu f_{\mu\nu}
+ i\theta\sigma^\mu\bar\theta\pa_\mu\zeta^\alpha
-{\textstyle{i\over 2}}\theta\theta\bar\sigma^{\lambda\dot{\alpha}\alpha}
\pa_\lambda\bar\zeta_{\dot{\alpha}}
-{\textstyle{i\over 2}}\theta\theta\bar\theta^{\dot{\gamma}}
\epsilon^{\beta\gamma}\sigma^\lambda_{\gamma\dot{\gamma}}
\sigma^\mu_{\beta\dot{\beta}}\bar\sigma^{\nu\dot{\beta}\alpha}\pa_\lambda
f_{\mu\nu}, 
\eeqa
and 
\beqa
\label{ecn4a}
\epsilon^{\mu\nu\sigma\lambda}\pa_\lambda f_{\rm{AR}\sigma\nu}
&\is&\eta^{\lambda\nu}\pa_\mu f_{\rm{SI}\lambda\nu} 
-2 \eta^{\mu\nu}\pa^\lambda f_{\rm{AI}\lambda\nu},\\
\label{ecn4b}
\pa_\mu v_\nu + \pa_\nu v_\mu &\is& -4 \eta_{\mu\nu}
\eta^{\sigma\lambda}f_{\rm{SR}\sigma\lambda},\\
\label{ecn4c}
\pa_\mu v_\nu -\pa_\nu v_\mu &\is& -8f_{\rm{AR}\mu\nu} 
-4\epsilon^{\sigma\lambda}_{\ \ \mu\nu}f_{\rm{AI}\sigma\lambda},\\
\label{ecn4d}
\pa^\lambda\pa_\lambda\zeta &\is& 0,
\eeqa
for symmetric and antisymmetric components of real and imaginary parts of 
$f_{\mu\nu}$ defined by
\beq
\label{econ5}
f_{\mu\nu} \equiv f_{\rm{SR}\mu\nu} + f_{\rm{AR}\mu\nu}
+ i\left(f_{\rm{SI}\mu\nu}+f_{\rm{AI}\mu\nu}\right) .
\eeq 
In fact, $\xi$ is chiral so that
in terms of $y^\mu \equiv x^\mu + i \theta\sigma^\mu\bar\theta$
it simplifies as
\beqa
\label{econ6}
\xi^\alpha &\is& \zeta^\alpha(y) +f_{\mu\nu}(y)\theta^\beta
\sigma^\mu_{\beta\dot{\alpha}}\bar{\sigma}^{\nu\dot{\alpha}\alpha}
+\theta\theta\kappa^\alpha(y),\\
&&\kappa^\alpha = -{\textstyle{i\over 2}}\bar\sigma^{\lambda\dot{\alpha}\alpha}
\pa_\lambda\bar{\zeta}_{\dot{\alpha}}.
\eeqa
Note that not all the components of $f_{\mu\nu}$ are uniquely
determined by the superconformal condition. Eq.(\ref{ecn4b}) leads to the 
usual conformal transformations and eq.(\ref{ecn4d}) is the well known
property of superconformal transformations, i.e. $\zeta$ is harmonic\cite{WZ}.

The detail of $F_\alpha^{\ \beta}$ is not important, 
but if we add for the sake of completeness, then
\beqa
\label{econ7}
 F_\alpha^{\ \beta} &\is& 
\sigma^\mu_{\alpha\dot{\alpha}}\bar\sigma^{\nu\dot{\alpha}\beta}f_{\mu\nu}
+ 2i \sigma^\mu_{\alpha\dot{\alpha}}\bar\theta^{\dot\alpha}\pa_\mu\zeta^\beta
-i\theta_\alpha\bar\sigma^{\lambda\dot{\alpha}\beta}
\pa_\lambda\bar\zeta_{\dot\alpha} 
-i\left(\theta_\alpha\bar\theta^{\dot\alpha}\epsilon^{\gamma\delta}
\sigma^\lambda_{\delta\dot{\alpha}} + \theta^\gamma\bar\theta^{\dot\alpha}
\sigma^\lambda_{\alpha\dot\alpha}\right) \sigma^\mu_{\gamma\dot\beta}
\bar\sigma^{\nu\dot\beta\beta}\pa_\lambda f_{\mu\nu}\nonumber\\ 
&\!\!\! &+ {\textstyle{1\over 2}}\theta^\gamma\bar\theta\bar\theta 
\epsilon^{\dot\alpha\dot\gamma}
\sigma^\mu_{\alpha\dot\alpha}\sigma^\lambda_{\gamma\dot\gamma}
\pa_\mu\pa_\lambda\zeta^\beta 
+ {\textstyle{1\over 2}}\theta\theta\bar\theta^{\dot\alpha}
\sigma^\mu_{\alpha\dot\alpha}\bar\sigma^{\lambda\dot\beta\beta}
\pa_\mu\pa_\lambda\bar\zeta_{\dot\beta} 
+{\textstyle{1\over 4}}\theta\theta\bar\theta\bar\theta
\sigma^\mu_{\alpha\dot\alpha}\bar\sigma^{\lambda\dot\alpha\delta} 
\sigma^\sigma_{\delta\dot\beta}\bar\sigma^{\nu\dot\beta\beta}
\pa_\mu\pa_\lambda f_{\sigma\nu} \quad\quad\quad
\eeqa 

Solving eqs.(\ref{ecn4a}-\ref{ecn4d}), we can obtain a representation of 
the generators for $N=1$ superconformal algebra as\footnote{We follow the
same normalization convention of \cite{WZ}.}
\beqa
\label{escf2}
M_{\mu\nu} &\is& x_\mu\pa_\nu -x_\nu\pa_\mu -\left(\theta\sigma_{\mu\nu}
\pth + \bar\theta\bar\sigma_{\mu\nu}\pthb\right),\\
P_\mu &\is& \pa_\mu,\\
D_d &\is& x^\mu\pa_\mu +\half\left(\theta\pth +\bar\theta\pthb\right),\\
R &\is& \theta\pth -\bar\theta\pthb,\\
K_\mu &\is& \left(2x_\mu x^\lambda -x^2\delta_\mu^{\ \lambda}
-\theta\theta\thetabar\thetabar\delta_\mu^{\ \lambda}\right)\pa_\lambda 
\nonumber\\
&&-\left(x^\lambda\sigma_\mu\bar\sigma_\lambda \theta +
2i\theta\sigma_\mu\thetabar\theta \right)\pth
-\left(x^\lambda\sigma_\mu\bar\sigma_\lambda \thetabar -
2i\theta\sigma_\mu\thetabar\thetabar \right)\pthb , \\
Q_\alpha &\is&{\pa\ \over\pa\theta^\alpha} - i\sigma^\lambda_{\alpha\dot\alpha}
\thetabar^{\dot\alpha}\pa_\lambda,\\
S_\alpha &\is& -\left(ix^\lambda\sigma_{\lambda\alpha\dot\beta}
\bar\sigma^{\mu\dot\beta\beta}\theta_\beta + \sigma^\mu_{\alpha\dot\beta}
\thetabar^{\dot\beta}\theta\theta\right)\pa_\mu 
-4i\theta_\alpha\theta\pth +\left(x^\lambda\sigma_{\lambda\alpha\dot\beta}
+2i\theta_\alpha\thetabar_{\dot\beta}
\right){\pa\ \over\pa\thetabar_{\dot\beta}}.
\eeqa

\newsection{Supersymmetric Dilaton Geometry}

For supersymmetry we consider vielbeins $e_\mu^{\ a}$ and their superpartners,
the gravitino, $\psi_\mu^{\ \alpha}$. Then the dilaton geometry of eq.(\ref{e4})
is given by
\beq
\label{es1}
e_\mu^{\ a} = \e^{\kappa\phi}\delta_\mu^{\ a}
\eeq
for $\phi$ defined in eq.(\ref{e4}) and 
\beq
\label{es2}
{\textstyle{1\over n}}\pa_\lambda v^\lambda\delta_\mu^{\ a} 
= \delta_\lambda^{\ a}\pa_\mu v^\lambda. 
\eeq
The supersymmetric dilaton geometry needs fermionic analog of eq.(\ref{es1}) 
and the following is a good candidate\footnote{We thank M. Grisaru for
suggesting the gamma trace.}:
\beq
\label{es3}
\psi_\mu^{\ \alpha} = \bar\sigma_\mu^{\dot\alpha\alpha}
\bar\psi_{\dot\alpha},
\eeq
where $\sigma$-matrices are those in flat space in the same spirit as 
eq.(\ref{es1}).

To see if eq.(\ref{es3}) works we need to show that its supersymmetric
Diff transformations correctly reproduces the superconformal transformations
of flat superspace. Note that 
under supersymmetric Diff, $\psi_\mu^{\ a}$ picks up an extra fermionic term
such that 
\beq
\label{es4}
\delta\psi_\mu^{\ \alpha} = v^\lambda\pa_\lambda\psi_\mu^{\ \alpha} +
\pa_\mu v^\lambda\psi_\lambda^{\ \alpha} + 2\pa_\mu\zeta^\alpha.
\eeq
Then for eq.(\ref{es3}) we obtain
\beq
\label{es4b}
\bar\sigma_\mu^{\dot\alpha\alpha}\delta\bar\psi_{\dot\alpha} 
= v^\lambda\bar\sigma_\mu^{\dot\alpha\alpha}\pa_\lambda\bar\psi_{\dot\alpha}
+\bar\sigma_\lambda^{\dot\alpha\alpha}\bar\psi_{\dot\alpha}\pa_\mu v^\lambda
+2\pa_\mu\zeta^\alpha.
\eeq
Contracting with flat $\sigma^\mu$ leads to 
\beq
\label{es5}
\delta\bar\psi_{\dot\alpha} = v^\mu\pa_\mu\bar\psi_{\dot\alpha}
+{\textstyle{1\over n}}\bar\psi_{\dot\alpha}\pa_\mu v^\mu
-{\textstyle{2\over n}}\sigma^\mu_{\alpha\dot\alpha}\pa_\mu
\zeta^\alpha.
\eeq
Consistency conditions between eqs.(\ref{es4b}-\ref{es5}) can be obtained
by plugging eq.(\ref{es5}) back into eq.(\ref{es4b}) and they read
\beq
\label{es6}
-{\textstyle{2\over n}}\bar\sigma_\mu^{\dot\alpha\alpha}
\sigma^\lambda_{\beta\dot\alpha}\pa_\lambda\zeta^\beta = 2\pa_\mu\zeta^\alpha. 
\eeq
This is the superconformal condition in flat space, the counter-part of 
eq.(\ref{es2}). Now it is easy to show that this is indeed equivalent to the 
well-known superconformal conditions. Solutions to eq.(\ref{es6}) can
be written as
\beq
\label{es7}
\pa_\mu\zeta^\alpha =
\bar\sigma_\mu^{\dot\alpha\alpha}\bar\eta_{\dot\alpha},
\eeq
and eq.(\ref{es6}) implies, for $n \neq 1$,
\beq
\label{es8}
2\bar\sigma_\mu^{\dot\alpha\alpha}\pa^\mu\bar\eta_{\dot\alpha}
=2\pa^\mu\pa_\mu\zeta^\alpha = {\textstyle{2\over n}}
\pa^\mu\pa_\mu\zeta^\alpha = 0.
\eeq
Therefore, $\zeta^\alpha$ is harmonic and $\bar\eta_{\dot\alpha}$ is a
constant spinor and that we have recovered the well known superconformal
conditions in flat superspace\cite{WZ}.

In the above sense, eqs.(\ref{es1},\ref{es3}) define the supersymmetric dilaton
geometry and we identify $\phi$ as the dilaton and $\psi$ as its superpartner,
the dilatino. (In practice, we shall later identify a scaled one as  the
dilatino. See the next section.) Their transformation rules under dilatations
are given by eq.(\ref{e5}) and eq.(\ref{es5}) respectively. 
Thus the supersymmetric dilaton geometry is an effective way of 
describing the supersymmetric dilatations in the flat space.
For the dilaton and dilatino to form a supermultiplet we must include the axion,
hence defining the dilaton-axion chiral supermultiplet, $(\phi_c,\psi)$.
The R-charge of the dilaton-axion multiplet can be easily determined from 
eq.(\ref{eax3}) and the superconformal symmetry such that its R-charge is zero.

\newsection{Effective Lagrangian}

\noindent
We demand the superconformally invariant lagrangian to be covariant in the
supersymmetric dilaton geometry. Thus, such a scale invariant
effective lagrangian for the dilaton and the dilatino can be read
off from the supergravity lagrangian\cite{rsugra}\cite{WB} using the metric of 
the dilaton geometry. For simplicity, let us consider the $N=1$ supergravity 
lagrangian in four dimensions
\beq
\label{el1}
\CL_{{\rm S.G.}} = -{\textstyle{1\over 2\kappa^2}}eR(e,\Omega)
+\half e \tilde\epsilon^{\lambda\sigma\mu\nu}\left(
\bar\psi_\lambda\bar{\tilde\sigma}_\sigma\CD_\mu\psi_\nu
-\psi_\lambda{\tilde\sigma}_\sigma\CD_\mu\bar\psi_\nu
\right),
\eeq
where $\tilde\sigma_\sigma$ and $\tilde\epsilon^{\lambda\sigma\mu\nu}$ 
are in curved space and 
$\CD_\mu = \pa_\mu +\half\Omega_\mu^{ab}\sigma_{ab}$. The spin
connection also contains the contribution from the gravitino. 
The axion kinetic energy term can be easily included as the 
antisymmetric tensor term, i.e. $H^2$ term.

In the dilaton geometry the spin connection is given by
\beqa
\label{el2}
&&\Omega_{\mu a}^{\ \ b} \equiv \omega_{\mu a}^{\ \ b} 
+ \kappa_{\mu a}^{\ \ b},\\
&&\quad\omega_{\mu a}^{\ \ b} = 
\kappa\left(\delta_\mu^{\ b}\delta_{\ a}^{\lambda}
-\eta^{b\lambda}\eta_{\mu a}\right)\pa_\lambda\phi ,
\nonumber\\
&&\quad\kappa_{\mu a}^{\ \ b} = -{\textstyle{\kappa^2\over 2}}
\e^{-\kappa\phi}\epsilon_{\mu a}^{\ \ bc}
\bar\psi\bar\sigma_c\psi
\nonumber
\eeqa
so that we obtain
\beq
\label{e13}
eR(e,\Omega) = 6\kappa^2 \e^{2\kappa\phi}\eta^{\mu\nu}\pa_\mu\phi\pa_\nu\phi
+3\kappa^4\bar\psi\bar\psi\psi\psi +\cdots,
\eeq
where the ellipsis is a total derivative. The gravitino term leads to
\beq
\label{e14}
\half e \tilde\epsilon^{\lambda\sigma\mu\nu}\left(
\bar\psi_\lambda\bar{\tilde\sigma}_\sigma\CD_\mu\psi_\nu
-\psi_\lambda{\tilde\sigma}_\sigma\CD_\mu\bar\psi_\nu
\right) = -3i\e^{\kappa\phi}\left(\bar\psi\bar\sigma^\mu\pa_\mu\psi
+ \psi\sigma^\mu\pa_\mu\bar\psi\right).
\eeq 
This is almost the dilatino kinetic energy term except the prefactor involving
the dilaton. This prefactor can be removed easily by scaling as
\beq
\label{e15}
\psi \equiv \e^{-\half\left(\kappa\phi+i{a\over f_a}\right)}\lambda,
\eeq
then the kinetic term leads to supersymmetric dilaton effective lagrangian
\beq
\label{e16}
\CL_{{\rm dil}} = -\half\e^{2\kappa\phi}\eta^{\mu\nu}\pa_\mu\phi\pa_\nu\phi
-\half\e^{2\kappa\phi}\eta^{\mu\nu}\pa_\mu a\pa_\nu a
-\half i \left(\bar\lambda\bar\sigma^\mu\CD_\mu\lambda
+ \lambda\sigma^\mu\CD_\mu\bar\lambda\right) 
-{\textstyle{\kappa^2\over 4}}\e^{-2\kappa\phi}
\bar\lambda\bar\lambda\lambda\lambda,
\eeq
where $\CD_\mu = \pa_\mu -{i\over 2f_a}\pa_\mu a$ and the dilatino
$\lambda$ is a Majorana-Weyl fermion. Note that the dilaton-dilatino
coupling shows that this is a four dimensional analog to the two-dimensional 
super-Liouville theory lagrangian (if the axion terms are dropped). 
In two dimensions the corresponding term is renormalizable, but
not in four dimensions. Perhaps, this could be the reason why this lagrangian
has never been investigated. Note that in this field configuration the prefactor
of the four-dilatino term is $\e^{-2\kappa\phi}$, not $\e^{+2\kappa\phi}$.

By construction, this effective lagrangian is invariant under superconformal
transformations: 
\beqa
\label{e17a}
\delta\e^{\kappa\phi} &\is& {\textstyle{1\over n}}\e^{\kappa\phi}D_\mu v^\mu,\\
\label{e17b}
\delta a &\is& v^\mu\pa_\mu a,\\
\label{e17c}
\delta\lambda &\is& {\textstyle{3\over 8}}\lambda\pa_\mu v^\mu +v^\mu\pa_\mu
\lambda 
-\half\e^{\half\left(\kappa\phi+i{a\over f_a}\right)}\sigma^\mu\pa_\mu\bar\zeta.
\eeqa 
Therefore, in particular, it is scale invariant.

There is another global supersymmetry that leaves the lagrangian invariant:
\beqa
\label{e18a}
\delta_\epsilon\phi &\is& 2(\epsilon\lambda + \bar\epsilon\bar\lambda), \\
\label{e18b}
\delta_\epsilon a &\is& -2i (\epsilon\lambda - \bar\epsilon\bar\lambda),\\
\label{e18c}
\delta_\epsilon\lambda 
&\is& \kappa(\epsilon\lambda + \bar\epsilon\bar\lambda)\lambda
+{\textstyle{1\over f_a}}(\epsilon\lambda - \bar\epsilon\bar\lambda)\lambda
+{\textstyle{5\over 8}}\kappa(\bar\epsilon\bar\lambda)\lambda 
+{\textstyle{i\over 4}}\e^{2\kappa\phi}\sigma^\mu\CD_\mu\phi\bar\epsilon.
\eeqa
As a matter of fact, this is not the only global supersymmetry we can obtain
but we have chosen it such that $\delta_\epsilon\phi$ does not involve $\phi$ 
in the RHS of eq.(\ref{e18a}). Since we have already got rid of auxiliary fields
implicitly, the proof of this global supersymmetry requires the equations of 
motion
\beqa
\label{e19a}
0 &\is& \pa_\mu\pa^\mu\phi + \kappa\pa_\mu\phi\pa^\mu\phi 
-\kappa\pa_\mu a\pa^\mu a +{\textstyle{\kappa^3\over 2}}
\e^{-4\kappa\phi}\bar\lambda\bar\lambda\lambda\lambda,\\
\label{e19b}
0 &\is& \pa^\mu\left(e^{2\kappa\phi}\pa_\mu a -{\textstyle{i\over 2f_a}}
\bar\lambda\bar\sigma_\mu\lambda\right),\\
\label{e19c}
0 &\is& i\bar\sigma^\mu\CD_\mu\lambda + {\textstyle{\kappa^2\over 4}}
\e^{-2\kappa\phi}\lambda\lambda\bar\lambda.
\eeqa

Next, let us consider couplings of chiral supermultiplets $(z^i,\psi^i)$
and vector supermultiplets $(A^a_\mu, \lambda^a)$.
The K\"ahler and super-potentials are now 
$K(\phi_c, \bar\phi_c, z^i, \bar{z}^i)$ 
and $W(\phi_c, z^i)$, where $z^i$ are now dilaton dressed fields 
and $\phi_c$ is given in eq.(\ref{eax1}). 
We choose $K = z^i\bar z^i$ (before dressing) so that the K\"ahler metric is 
$g_{ij^*} = \delta_{ij}$.
Then the 
matter part is given by
\beq
\label{e20}
\CL_{{\rm matter}}= \CL_1 +\CL_2 -V,
\eeq
where
\beqa
\label{e21a}
\CL_1 &\is&
-\CD^{(1)}_\mu z^i \bar\CD^{(1)\mu}\bar{z}^i
-i\bar\psi^i\bar\sigma^\mu\tilde\CD_\mu\psi^i
-{\textstyle{1\over 16\pi}}\tau^I F^{a}_{\mu\nu}F^{a\mu\nu}
+{\textstyle{1\over 32\pi}}
\tau^R\epsilon^{\mu\nu\lambda\sigma}F^{a}_{\mu\nu}F^{a}_{\lambda\sigma}
\nonumber\\
&& -i\bar\lambda^a\bar\sigma^\mu\tilde\CD_\mu\lambda^a
+\left(i\sqrt{2} g\e^{-i{2\over f_a} a}T^a_{ij}z^j\psi^i\lambda^a 
+ {\rm h.c.}\right)
\\
\label{e21b}
\CL_2 &\is& 
\kappa\e^{-\kappa\phi}\left(2g\e^{i{2\over f_a}a} 
D^{(a)}\bar\lambda\bar\lambda^a
-\sqrt{2}\bar\CD^{(1)}_\mu\bar{z}^i\psi^i
\sigma^\mu\bar\lambda +{\rm h.c.}\right) \nonumber\\
&& + \kappa^2\e^{-2\kappa\phi}\Big[{\textstyle{1\over 2}}
(\bar\psi^i\bar\lambda)(\psi^i\lambda)
+{\textstyle{i\over 4}}\left(\bar{z}^i\CD^{(1)}_\mu
z^i -{z}^i\bar\CD^{(1)}_\mu\bar{z}^i\right) \bar{\psi}^j\bar\sigma^\mu\psi^j
-{\textstyle{3\over 2}}(\bar\lambda^a\bar\lambda)(\lambda^a\lambda) \nonumber\\
&& \qquad +{\textstyle{i\over 4}}\left(\bar{z}^i\CD^{(1)}_\mu z^i 
-{z}^i\bar\CD^{(1)}_\mu\bar{z}^i\right) \bar{\lambda}^a\bar\sigma^\mu\lambda^a
-{\textstyle{1\over 16\pi}}\tau^I (\bar\psi^i\bar\lambda^a)(\psi^i\lambda^a)
+{\textstyle{3\over 8}}(\bar\lambda^a\bar\lambda^b)(\lambda^a\lambda^b)\Big]
\nonumber\\
&& -{\rm exp}\left({\half\kappa^2 z^k\bar{z}^k \e^{-2\kappa\phi}}\right)
\Big[6\kappa^2\e^{-2\kappa\phi-i{4\over f_a} a}
\bar{W}\bar\lambda\bar\lambda
-i2\sqrt{2}\kappa\e^{-\kappa\phi-i{4\over f_a} a}D_i W\psi^i\lambda
\nonumber\\
&& \qquad +\half\e^{-i{4\over f_a} a}\psi^i\psi^j\left(\pa_i\pa_j W 
+2\kappa^2\e^{-2\kappa\phi}\bar{z}_iD_jW
-\kappa^4\e^{-4\kappa\phi}\bar{z}_i\bar{z}_j W
\right) +{\rm h.c.}\Big]\ \ 
\\
\label{e21c}
V &\is& V_F + V_D,\\
&&V_F\equiv
{\rm exp}\left({\kappa^2 z^i\bar{z}^i \e^{-2\kappa\phi}}\right)
\left(D_jW \bar{D}_j\bar{W} -3\kappa^2\e^{-2\kappa\phi}W\bar{W}\right),
\nonumber\\
&&V_D\equiv\half g^2 D^{(a)} D^{(a)}. \nonumber
\eeqa
In the above, we used
$\tau \equiv {\theta\over 2\pi} + i {4\pi\over g^2} $,
$D^{(a)}\equiv \bar z^i T^a_{ij} z^j $, 
$\CD^{(k)}_\mu \equiv \pa_\mu - k(\kappa \pa_\mu\phi +{i\over f_a}\pa_\mu a)
 +igA^a_\mu T^a $, $\tilde\CD_\mu\equiv\CD^{(0)}_\mu-i{3\over 2f_a}\pa_\mu a $, 
$\pa_i\equiv \pa/\pa z^i$ and 
$D_iW\equiv \pa_iW +\kappa^2\e^{-2\kappa\phi}\bar{z}_i W $.
Note that in the fermionic kinetic energy terms
there are no dilaton contributions.
This is because the spin connection contribution from the dilaton geometry
exactly cancels the term from the dilaton dressing.
This lagrangian is superconformally invariant classically by construction 
due to the supergravity of the supersymmetric dilaton geometry. 
The superconformal transformation rules for matter fields are the standard ones
and straightforward. 

The full lagrangian also has a global supersymmetry
\beqa
\delta_\epsilon\phi &\is& 2(\epsilon\lambda +\bar\epsilon\bar\lambda), \\
\delta_\epsilon a &\is& -2i(\epsilon\lambda -\bar\epsilon\bar\lambda), \\
\delta_\epsilon\lambda &\is& \half\left(\kappa\delta_\epsilon\phi 
+ {\textstyle{i\over f_a}}\delta_\epsilon a\right)\lambda
+{\textstyle{5\over 8}}\kappa(\bar\epsilon\bar\lambda)\lambda 
+{\textstyle{i\over 4}}\e^{2\kappa\phi}\sigma^\mu\CD_\mu\phi\bar\epsilon
+{\textstyle{i\over 16}}\left(\bar{z}^i\sigma^\mu\CD^{(1)}_\mu z^i
-{z}^i\sigma^\mu\bar\CD^{(1)}_\mu \bar{z}^i\right)\bar\epsilon \nonumber\\
&& +{\textstyle{3\over 8}}\kappa(\bar\epsilon\bar\psi^i)\psi^i
-{\textstyle{5\over 8}}\kappa(\bar\epsilon\bar\lambda^a)\lambda^a
+{\textstyle{i\over 2\sqrt{2}}}\kappa^2\e^{-\kappa\phi}
\left(z^i\bar\epsilon\bar\psi^i -\bar{z}^i\epsilon\psi^i\right)\lambda
+{\rm exp}
\left(\half\kappa^2 z^i\bar{z}^i\e^{-2\kappa\phi}\right)\bar{W}\epsilon ,
\quad \qquad\\
\delta_\epsilon z^i &\is& i\sqrt{2}\epsilon\psi^i\e^{\kappa\phi} 
+\left(\kappa\delta_\epsilon\phi + {\textstyle{i\over f_a}}
\delta_\epsilon a\right)z^i, \\
\delta_\epsilon\psi^i &\is& {\textstyle{3\over 2}}
\left(\kappa\delta_\epsilon\phi 
+ {\textstyle{i\over f_a}}\delta_\epsilon a\right)\psi^i
-\kappa(\bar\epsilon\bar\lambda)\psi^i
+\sqrt{2}\e^{\kappa\phi+i{a\over f_a}}\sigma^\mu\CD^{(1)}_\mu z^i\bar\epsilon 
+{\textstyle{i\over 2\sqrt{2}}}\kappa^2\e^{-\kappa\phi-i{a\over f_a}}
\left(z^j\bar\epsilon\bar\psi^j -\bar{z}^j\epsilon\psi^j\right)\psi^i
\nonumber\\
&& -i\sqrt{2}{\rm exp}
\left(\half\kappa^2 z^k\bar{z}^k\e^{-2\kappa\phi}\right)
\e^{\kappa\phi+i4{a\over f_a}}\bar{D}^i\bar{W}\epsilon ,\\
\delta_\epsilon A_\mu^a &\is& -\e^{\kappa\phi}
\left(\e^{i{2\over f_a} a}\epsilon\sigma_\mu\bar\lambda^a
-\e^{-i{2\over f_a} a}\bar\epsilon\bar\sigma_\mu\lambda^a\right),\\
\delta_\epsilon\lambda^a &\is& i\e^{\kappa\phi}\e^{i{2\over f_a} a}
\left(F^a_{\mu\nu}\sigma^{\mu\nu}-igD^{(a)}\right)\epsilon
+{\textstyle{3\over 2}}\left(\kappa\delta_\epsilon\phi 
+ {\textstyle{i\over f_a}}\delta_\epsilon a\right)\lambda^a
-{\textstyle{i\over 2\sqrt{2}}}\kappa^2\e^{-\kappa\phi}
\left(z^i\bar\epsilon\bar\psi^i +\bar{z}^i\epsilon\psi^i\right)\lambda^a.
\eeqa

The dilaton, axion and dilatino form additional chiral supermultiplet.
The dilaton and axion couplings are unavoidably nonrenormalizable. 
As $\kappa\to 0$ and $f_a\to\infty$, 
$\CL_2\to -\half\psi^i\psi^j\pa_i\pa_j W $
and $\CL_{{\rm matter}}$ reduces to the $N=1$ supersymmetric gauge theory.
Thus we can regard $\CL_{{\rm dil}} +\CL_{{\rm matter}}$ as the scale 
invariant generalization of $N=1$ supersymmetric gauge theory incorporating 
the dilaton chiral supermultiplet. All the nonrenormalizable terms are 
dictated by the supersymmetric dilaton geometry.

To be consistent with the global supersymmetry the superpotential $W$ must
lead to a positive semidefinite scalar potential $V$ for a superconformally
invariant vacuum. Thus $W$ cannot be arbitrary, but it is not difficult to 
construct examples. In general, however, it is not 
necessary to satisfy $V_F=|\pa_a \tilde{W}|^2$ 
for some superpotential ($a$ now
runs over the dilaton too) because there could be soft breaking terms due
to the dilaton much analogous to the supergravity. It is known in the
supergravity that the spontaneous breaking of local supersymmetry manifests
itself as soft breaking of global supersymmetry\cite{barbieri}.
In the case of unbroken supersymmetry with broken superconformal symmetry,
the effective scalar potential might be expressed as  $V_F=|\pa_a \tilde{W}|^2$,
where $\tilde{W}$ is not the same as $W$ for an obvious reason.
A similar line of thought also appears in softly broken $N=2$ supersymmetric
QCD\cite{alvarez}.
I shall discuss more details in the following section.

\newsection{Symmetry Breaking}

\noindent
In the supersymmetric case breaking scale symmetry is nothing but breaking
the superconformal symmetry so that it automatically addresses the issue of 
supersymmetry breaking. 
The conventional wisdom in the absence of the dilaton is that the scale
and R symmetries are anomalous, yet these anomalies leave
the Poincar\'e supersymmetry unbroken.

In the presence of the dilaton we presume that the scale and R symmetries
should be spontaneously broken. Thus we need a nonanomalous dilaton sector
in which scale symmetry is spontaneously broken. In fact, this can also be
done at nontrivial fixed points, where $\beta$-functions vanish, so that
we can prevent the trace anomaly from causing any complications.
The violation of the conservation of the scale current consists of 
two terms: the dilaton mass term and the trace anomaly
This is much analogous to the pion case of PCAC, in which the conservation 
of the axial current is violated by two terms: pion mass term and the axial
anomaly.
Looking at the superconformal algebra, we can
easily understand that, if scale symmetry is spontaneously broken,  the
Poincar\'e supersymmetry could also be broken, unless the R-symmetry breaking
is related to the scale symmetry breaking in a specific way.
Furthermore, explicit soft supersymmetry breaking can occur too. 
This makes the situation much more complicated than the bosonic case.

Let us first consider the consequence of unbroken superconformal symmetry, 
hence, unbroken Poincar\'e supersymmetry. 
In this case, for the Lorentz invariant vacuum,
eq.(\ref{escf1d}) implies 
\beq
\label{e22}
\langle 0|D_d|0\rangle ={\textstyle{3\over 2}}i\langle 0|\gamma_5 R|0\rangle.
\eeq
and in particular the vacuum is also an eigenstate for both $D_d$ and $R$.
Due to the presence of $\gamma_5$, for nonvanishing vacuum expectation values, 
this identity can only be  satisfied if the vacuum is fermionic such that
\beq
\label{e23}
|0\rangle ={|+\rangle \choose |-\rangle}.
\eeq
The vacuum states are eigenstates of $D_d$ or $R$, although other states are
not. In particular, $|0\rangle$ need to be a Majorana spinor so that
\beq
\label{e24}
R|\pm\rangle = \pm c|\pm\rangle.
\eeq
Then for unbroken supersymmetry
\beq
\label{e25}
D_d|\pm\rangle = {\textstyle{3\over 2}}c|\pm\rangle.
\eeq
Hence, for unbroken Poincar\'e supersymmetry the $R$-charge of the vacuum must
be specifically related to its scale dimension\cite{sohnius}.  Since the 
invariant vacuum should not carry a scale-dimension, we can consistently 
choose $c=0$.
Next, for a bosonic vacuum, eq.(\ref{e22}) is true only if
\beq
\label{e26}
\langle 0|D_d|0\rangle = 0 = \langle 0| R|0\rangle.
\eeq
The bosonic supersymmetric vacuum cannot carry a scale-dimension or R-charge.
Therefore, for superconformally invariant vacuum we can always choose 
$D_d|0\rangle = 0 = R|0\rangle$. 

This suggests us that we can use $D_d|0\rangle \neq 0$ 
($R|0\rangle\neq 0$) as an indication for spontaneous breaking of scale 
symmetry (R-symmetry, respectively), hence, 
spontaneous breaking of conformal symmetry. This is nothing unusual for the 
R-symmetry, which is, more or less, internal.
For the scale symmetry this is true only if it is not realized in the 
Wigner-Weyl way. For example, the translational symmetry is not broken
even if the vacuum is not translationally invariant, which is the case
of nonvanishing vacuum energy. This is because one can still find a unitary 
operator that leaves the Hilbert space invariant under the translation. 
So we have no violation of energy-momentum
conservation despite the nonvanishing vacuum energy.
Without the dilaton, equally one can argue this is the case. But, since the 
dilaton does not transform as a quasiprimary field under dilatations, the scale
symmetry is not realized in the Wigner-Weyl way in the presence of the 
dilaton. Therefore, we can use $D_d|0\rangle \neq 0$ as the symmetry breaking
condition. One can easily check that this is also consistent with
the conservation law of the scale current much the same way as in any other
internal symmetry cases.

If supersymmetry is broken spontaneously, the vacuum states are no longer 
eigenstates of $D_d$ because $H$ and $D_d$ do not commute (this is the case for 
soft breaking too). For
$E_0 =\langle 0|H|0\rangle \neq 0$ the vacuum seems to violate
$[H,D_d] = iH$ because $\langle 0|[H, D_d]|0\rangle = 0$.
In fact, this happens for any state with nonzero energy eigenvalue if
we define $H|E\rangle =E|E\rangle$ with $\langle E|E\rangle = 1$ naively.
This normalization condition is true only if the spectrum is discrete. 
The matter of the fact is that
the conformal symmetry demands the spectrum to be continuous\footnote{The author
thanks D.Z. Freedman for pointing out this.}, if $E\neq 0$.
Therefore, the naive normalization condition is not correct, but should be
replaced by $\langle E|E'\rangle = \delta(E-E')$.
In particular, for continuous spectrum let $E_a = \e^aE_0$, $a\geq 0$, then
$\langle E_b|E_a\rangle = {1\over E_a}\delta(b-a)$.
Now $E_0 =\int dE_b\langle E_b|H|0\rangle 
= -i\int dE_b\langle E_b|[H,D_d]|0\rangle$, provided that 
$\langle E_b|D_d|E_a\rangle = -i{1\over\sqrt{E_bE_a}}
{\pa\over\pa b}\delta (b-a)$.
In the superconformal case, $[R,H] = 0 =[D_d,R]$ implies the energy eigenstate
can be written as $|E,r\rangle = \sum_d a_d(E) |d,r\rangle$, where
$D_d|d,r\rangle =d|d,r\rangle$. 
Then 
\beq
\label{e27}
\langle E_b|D_d|E_a\rangle =\sum_d d a_d^*(E_b) a_d(E_a) 
= -i{1\over\sqrt{E_bE_a}}{\pa\over\pa b}\delta (b-a).
\eeq
The normalization condition in turn requires 
\beq
\label{e28}
\sum_d a_d^*(E_b)a_d(E_a) ={1\over E_a}\delta(b-a)
\eeq
To be consistent, such $a_d(E)$ must exist in superconformal field theories and
the above indeed yield a solution
\beq
\label{e29}
a_d(E_a) = {1\over \sqrt{2\pi E_a}}\e^{-iad}.
\eeq
Then, the continuous spectrum can be constructed in terms of
$|E_a\rangle = \e^{-\half a}\e^{-iaD}|E_0\rangle$ for any constant $a\geq 0$.
In nonsupersymmetric case, despite the absence of $[R,D_d]=0$, a similar
construction is possible as long as the spectrum is positive 
definite.
Thus energy eigenstates of nonvanishing energy eigenvalues
are superpositions of dilatation eigenstates. 

Notice that we can in fact construct states $|E_a\rangle$ for $a<0$.
Recall that in the above $a\geq 0$ condition is required simply due to the 
assumption that $|E_0\rangle$ is the vacuum $|0\rangle$. This implies if
the conformal symmetry is unbroken, the only possible vacuum has to be 
$E_0 = 0$ to be consistent, and that $D_d|0\rangle = 0$.
Hence, if supersymmetry is broken with nonzero vacuum energy, the conformal 
symmetry must be broken at the same time to make sure 
the states below the vacuum decoupled.

Next, let us ask if the superconformal symmetry can be broken without breaking 
the Poincar\'e supersymmetry. If $Q|0\rangle = 0$ and $D_d|0\rangle\neq 0$
(or $R|0\rangle\neq 0$) can be consistently satisfied, this can happen. 
For unbroken supersymmetry the vacuum must carry
a specific R-charge and scaling dimension. 
A supersymmetric vacuum is Poincar\'e invariant so that the superconformal 
algebra forces the vacuum to be a zero mode of $S$.
Therefore, $S|0\rangle = 0$ implies $R|0\rangle\neq 0$ and the vacuum must be 
fermionic such that $(D_d-i(3/2)\gamma_5 R )|0\rangle = 0$.
Thus R-symmetry is also broken and that the scale symmetry breaking is
specifically related to the R-symmetry breaking.

This indicates that, if the scale symmetry is spontaneously broken in the
bosonic vacuum, the Poincar\'e supersymmetry must be broken as
well as  the superconformal symmetry.  This can be a new way of breaking the
supersymmetry using the dilaton. It implies that all of their
symmetry breaking scales must be the same to be consistent. This puts a very
severe constraint on the possibility of spontaneous superconformal symmetry
breaking.

To achieve spontaneous breaking of the scale symmetry in the dilaton sector,
we first need to take care of the issue of the trace anomaly.
If there is a trace anomaly in the dilaton sector, then the scale symmetry is
explicitly broken and the existence of the dilaton is meaningless.
The pure dilaton sector only allows a specific form of potential which is
unavoidably nonrenormalizable to be conformally invariant.
Such a potential always puts the vacuum at infinity.

The remedy in nonsupersymmetric case is,
as noted in \cite{mydil}, introducing a dynamical scale which transforms under
dilatations as
\beq
\label{e30}
\delta M = \alpha M,
\eeq
we can in fact write down a dilaton effective potential which is free from
a trace anomaly: 
\beq
\label{e31}
V_{\phi,{\rm eff}} = {\Lambda\over 4}\e^{4\kappa\phi} +
{1\over 64\pi^2}\left(4\kappa^2\Lambda\e^{2\kappa\phi}\right)^2
\left(\log{4\kappa^2\Lambda\e^{2\kappa\phi}\over M^2} 
-{\textstyle{3\over 2}}\right).
\eeq
One can easily check that this effective potential is scale invariant
incorporating eq.(\ref{e30}). Without eq.(\ref{e30}), the 
$\phi\e^{4\kappa\phi}$ term is not scale invariant.
Note that introducing such a dynamical scale does not change trace anomalies
in other sectors because it does not modify properties of $\beta$-functions.
$M$ can be simply used as the scale which enters in the renormalization group
equation. Also there is no spontaneous breaking of scale symmetry in other
sectors. The basic difference lies on the fact that the dilaton is not a
quasiprimary field, whilst all other fields are quasiprimary fields.

This effective dilaton potential indeed admits a new vacuum in which the dilaton
gets a nontrivial vacuum expectation value and becomes massive. 
The new minimum located at
\beq
\label{e32}
\langle\phi\rangle = {1\over 2\kappa}\left(1 - {\pi^2\over \kappa^4\Lambda}
+\log{M^2\over 4\kappa^2\Lambda}\right)
\eeq
for any $\kappa^4\Lambda$.
This new vacuum is no longer invariant under dilatations so that the scale
symmetry is  spontaneously broken. The conservation law of the total dilatation
current in general will be violated by two terms as in the PCAC case of the
pions:
\beq
\label{e33}
\pa_\mu S^\mu = -{m_\phi^2\over \kappa} \phi + {{\rm trace\ anomaly}}
\eeq
for small $\phi$, where for the potential eq.(\ref{e31})
the dilaton mass is given by
\beq
\label{e34}
m^2_\phi = {\kappa^2 M^4\over 8\pi^2}\,
{\rm exp}\left\{2\left(1-{\pi^2\over \kappa^4\Lambda}\right)\right\}.
\eeq
Note that this dilaton mass is precisely the one arises in  $V_{\phi,{\rm
eff}}$ after shifting the vacuum as $\phi\to \phi + \langle\phi\rangle$, hence  
confirming that  our notion of spontaneous scale symmetry breaking is
consistent.
Once we obtain a vacuum with broken scale symmetry, $M$ can be fixed to a
numerical value. Thus, the dilaton mass depends on still-undetermined
three parameters. 

Eq.(\ref{e31}) cannot be obtained in the supersymmetric case because the 
corresponding superpotential must contain $\sqrt{\phi}$ term to lead to the
given scalar potential. Nevertheless, the same idea can be applied and
superpotentials that lead to spontaneous breaking of scale symmetry can be 
constructed. A useful form of the pure dilaton part of a scalar potential which 
exhibits spontaneous breaking of the scale symmetry in the same way as the 
bosonic case is
\beq
\label{e35}
V_{{\rm dil, eff}} = \kappa^2\Lambda\e^{4\kappa\phi}(\phi - a_1)(\phi -a_2),
\eeq
where
\beq
\label{e36}
\kappa a_i = \half\ln{\kappa^2\Lambda\over M^2} +c_i, \ i=1,2,
\eeq
and $c_i$'s are constants independent of $M$. The specific forms of $a_i$
are dictated to make the potential scale invariant so that there is no
trace anomaly from this potential.
This scalar potential can be derived from a superpotential, for example,
with one scalar field, $z$, other than the dilaton,
\beq
\label{e37}
W_{{\rm eff}}(\phi_c, z) = 
{\sqrt{\Lambda}\over 2\sqrt{3}}(a_1-a_2)\e^{3\kappa\phi_c} + 
\sqrt{\Lambda}\kappa z\e^{2\kappa\phi_c}\left(\phi_c -\half(a_1+a_2)\right).
\eeq
The first term is an R-symmetry breaking term that vanishes as $(a_1-a_2)\to 0$.
$V_{{\rm dil,eff}}$ has a scale symmetry breaking vacuum at 
\beq
\label{e38}
\langle\phi\rangle = v \equiv \half\left(a_1+a_2 -{1\over 2\kappa} 
+\sqrt{(a_1-a_2)^2 + {1\over 4\kappa^2}}\right) 
\eeq
with $V_{{\rm dil,eff}}(\langle\phi\rangle)\leq 0$. 
The equality is only for $a_1 = a_2$.
The dilaton mass can be easily computed after shifting the vacuum and reads
\beq
\label{e39}
m_\phi^2 = 2\kappa^2\Lambda\e^{4\kappa v}\left(1 + 2\kappa(2v -a_1 -a_2)\right).
\eeq
As $v\to -\infty$, $m_\phi\to 0$, confirming the dilaton is massless in the
scale invariant asymptotic vacuum.

To be more precise, the vacuum must be a vacuum for $V$, not just the pure
dilaton part $V_{{\rm dil}}$. The vacuum structure of $V$ is fairly complicated
even with just one chiral multiplet. It includes vacua for both
$V = 0$ and $V \neq 0$. For one chiral multiplet the vacuum we obtained for
$V_{{\rm dil}}$ is not stable along the $z$ direction if $a_1 \neq a_2$. 
Since the potential is bounded below, there must be another vacuum nearby that 
satisfies $V_F < 0$ and ${\pa V\over \pa z} = 0 = {\pa V\over \pa \phi_c}$. 
However, we expect either the scale symmetry must be also explicitly broken so 
that states with negative vacuum states can be allowed with potential bounded 
below, or $V= V_F + V_D \geq 0$ due to the D-term contribution.
If not, this is not a desirable result because we want $V\geq 0$.
Hence we may need to include more than one chiral multiplet.

If $a_1 = a_2 \equiv v$, then the vacuum is stable along the $z$ direction too.
In this case the vacuum energy also vanishes. One can also easily check that
the supersymmetry is unbroken in this vacuum, although the scale symmetry is
spontaneously broken with $\langle \phi\rangle = v$. The dilaton mass
in this case is $m_\phi^2 = 2\kappa^2\Lambda\e^{4\kappa v}$. If we assume
$c_i = 0$, then $m_\phi^2 = 2\kappa^4\Lambda^2/M^2$. 
This is an example that the scale symmetry breaking does not lead to 
broken supersymmetry as we analyzed before using the algebraic structure.

If $a_1 = a_2$, there is no axion contribution to the scalar potential
in my examples. Otherwise, however, 
the scalar potential explicitly contains the axion potential in some cases.
The axion potential roughly takes the form of
$f(\phi, z, \bar{z}) (a_1-a_2)(g(\phi)z\cos{a\over f_a}
+\bar{z}{a\over f_a}\sin{a\over f_a})$ for smooth functions $f$ and $g$.
In particular, the potential is not periodic for the axion if $a\neq 0$. 
Anyhow, it still yields extrema at $a = 0$. 
This nonperiodicity for $a_1\neq a_2$ is because the corresponding term
in the superpotential does not have R-symmetry. Recall that the dilaton 
multiplet has R-charge zero and $z$ has R-charge two. 
All the vacua I obtain determine $\langle a\rangle =0$ if $a_1\neq a_2$.

Next, let us ask if there is a vacuum with a vanishing vacuum energy and broken
supersymmetry. For one chiral multiplet case we can find a solution, taking
advantage of the Polonyi solution\cite{jpol}. 
If we assume 
\beq
\label{e40}
{1\over 2\sqrt{3}}{a_1-a_2\over \phi_c -\half(a_1 + a_2)}\e^{\kappa\phi_c}
= \pm (2-\sqrt{3})\e^{\kappa\phi},
\eeq
the wanted solution is
\beqa
\label{e41}
\kappa\langle\phi\rangle &\is& \half (a_1 + a_2) +2, \ \langle a\rangle = 0, \  
\kappa(a_1-a_2) = \pm 4\sqrt{3}(2-\sqrt{3}),\\
\label{e42}
\kappa \langle z\rangle &\is& \pm (\sqrt{3} -1)\e^{\kappa\langle\phi\rangle},
\eeqa
and the supersymmetry is broken, most likely, softly. The unusual property
for these two Polonyi vacua is that the conservation law of the scale current
is now violated by the dilaton term as well as a constant term.
The good news is this constant term is related to the dilaton mass so that
it still vanishes as the dilaton mass goes to zero. In this sense it still
satisfies the notion of PCDC. If there is nonvanishing D-term contribution, the
vacuum energy is no longer vanishing.

With one chiral multiplet we are not able to obtain an explicit vacuum
solution for $V> 0$. This, however, will not be a global minimum anyway
so that it is less interesting.
The asymptotic vacuum at $z = 0$ and $\phi\to -\infty$, where both
supersymmetry and scale symmetry are unbroken, is surrounded by a
valley because $V\neq 0$ if $z\neq 0$ as $\phi\to -\infty$. 
All vacua with $V=0$ are degenerate with the asymptotic vacuum so that there
could be nonperturbative states associated.
We expect there are a lot more interesting structures hidden in the case with
more than one chiral multiplet, and we will report the results elsewhere in
the near future.

\newsection{Conclusions}

\noindent
I have defined the scale symmetry in the flat spacetime using the 
diffeomorphism structure of the supersymmetric dilaton geometry. It naturally
incorporates the dilaton-axion multiplet in the superconformal generalization
of supersymmetric gauge theories. Explicit examples of spontaneous breaking of
the scale symmetry have been demonstrated. Depending on the details of the
dilaton structure, both vacua with or without the Poincare supersymmetry can be
obtained.

I have only analyzed the cases that allow asymptotic superconformally invariant
vacuum. However, other possibilities can also be speculated. For example, if 
there is a mechanism to lead to the dilatino condensation, then such an
asymptotic vacuum can be completely destabilized so that the vacua with a
vanishing vacuum energy we obtained will not run away toward
the asymptotic vacuum. 

One disadvantage (or perhaps, it might turn out to be a new discovery.) is, if
we demand the low energy scale symmetry is the same as the scale symmetry in
gravity, the Diff symmetry of curved spacetime must appear as spontaneously
broken down to the volume-preserving diffeomorphism (SDiff) symmetry, which 
nevertheless still contains the Poincar\'e symmetry.  Although this is nothing
against any current experimental observations, some readers may find it very
difficult to accept it. We hope nature herself will clarify this in future.

$N=2$ generalization is expected to address the results obtained in
Seiberg-Witten model\cite{rSW} in the line of \cite{alvarez}. In the  $N=1$
examples given the dilaton-axion enters holomorphically and the details of the
dilaton-axion produces both broken and unbroken supersymmetry. Thus we may be
able to reproduce similar results in this approach.
If there is a way to incorporate the gauge coupling constant $\tau$ with
$\phi_c$, it is consistent with the SW model. In my approach, this is impossible
if the theory is defined in four dimensions alone. However, compactifications
from a higher dimensional construction would allow the identification of  
$\phi_c = \tau$ because the phenomenon of no direct coupling between gauge 
fields and the dilaton is unique to four dimensions.
If this turns out to be true, then we can identify the Peccei-Quinn symmetry
as the R-symmetry in the supersymmetric scale symmetry. Furthermore, we might
be able to obtain QCD from the SW model, explicitly realizing the line of idea
as in \cite{alvarez}.
I hope to report the progress in the near future.

\bigskip\bigskip
\noindent
{\large\bf{Acknowledgements:}} I thank R. Arnowitt, 
D.Z. Freedman, M. Grisaru and G. Mack for helpful correspondence.

\bigskip\bigskip\bigskip
\appendix{\noindent\large\bf Appendix}\hskip-3mm
\newsection{Supersymmetric Volume-Preserving Diffeomorphisms}
\vskip-.5in

\noindent
The volume-preserving diffeomorphisms are defined by diffeomorphisms 
that leave a volume element invariant. This can be defined for any manifolds
with or without a boundary. The main advantage of decomposing diffeomorphisms
into volume-preserving diffeomorphisms and the rest is because the rest
are in fact conformal diffeomorphisms.

In the bosonic case the defining condition is
\beq
\label{eap0}
\delta\sqrt{g} = \half g^{\mu\nu}\delta g_{\mu\nu} = \nabla_\mu v^\mu = 0.
\eeq
In the presence of a boundary $v^\mu$ has to be also parallel with the
boundary. This is the precise condition none of conformal diffeomorphisms
can satisfy which requires $\nabla_\mu v^\mu \neq 0$.
Thus the dilaton geometry excludes volume-preserving diffeomorphisms to deal
only with the conformal diffeomorphisms.

To define the supersymmetric dilaton geometry we need to define the
supersymmetric generalization of SDiff. As ordinary SDiff transformations leave
the volume density $\sqrt{g}$  invariant, supersymmetric SDiff transformations
should leave the chiral density $\CE$ invariant, hence $\psi$ also invariant. 
In curved spacetime, therefore, we also demand
$\delta(\sigma^\mu_{\alpha\dot\alpha}\psi_\mu^\alpha) = 0$. 
Using $\sigma^\mu_{\alpha\dot\alpha}\psi_\mu^\alpha = E_a^{\
\mu}\sigma^a_{\alpha\dot\alpha}\psi_\mu^\alpha$, we obtain
\beq
\label{eap1}
\delta\bar\psi_{\dot\alpha} = v^\mu\pa_\mu\bar\psi_{\dot\alpha} 
-{\textstyle{2\over n}}\sigma^\mu_{\alpha\dot\alpha}\pa_\mu\zeta^\alpha = 0.
\eeq
Thus, together with eq.(\ref{eap0}), this defines
the supersymmetric SDiff.

%\nopagebreak

{\renewcommand{\Large}{\large}

}

\end{document}